\documentclass[aps,prl,twocolumn,showpacs,superscriptaddress,groupedaddress]{revtex4}
\usepackage{mathrsfs}
\usepackage{graphicx}
\usepackage{bm}        
\usepackage{amssymb}   
\usepackage{amsmath}
\usepackage{color}

\def\bea{\begin{eqnarray}}
\def\eea{\end{eqnarray}}

\begin{document}


\title{Elastic Model for Dinucleosome Structure and Energy}


\author{Hashem Fatemi}%
\thanks{H.F. and F.K. contributed equally to this work.}
\affiliation{Department of Physics, Institute for Advanced Studies in
Basic Sciences (IASBS), Zanjan 45137-66731, Iran}

\author{Fatemeh Khodabandeh}%
\thanks{H.F. and F.K. contributed equally to this work.}
\affiliation{Department of Physics, Institute for Advanced Studies in
Basic Sciences (IASBS), Zanjan 45137-66731, Iran}

\thanks{H.F. and F.K. contributed equally to this work.}

\author{Farshid Mohammad-Rafiee}%
\email{farshid@iasbs.ac.ir}
\affiliation{Department of Physics, Institute for Advanced Studies in
Basic Sciences (IASBS), Zanjan 45137-66731, Iran}

\date{\today}

\keywords{$^*$ F.K. and H.F. contributed equally to this work.}

\begin{abstract}
The equilibrium structure of a Dinucleosome is studied using an elastic model that takes into account the force and torque balance conditions. Using the proper boundary conditions, it is found that the conformational energy of the problem does not depend on the length of the linker DNA. In addition it is shown that the two histone octamers are almost perpendicular to each other and the linker DNA in short lengths is almost straight. These findings could shed some light on the role of DNA elasticity in the chromatin structure.

\end{abstract}

\maketitle

\section*{I. Introduction}

Packaging of genomic DNA into chromatin is essential for eukaryotic cells. The nucleosomes, which are the building blocks of the chromatin, are connected to each other with 10-90 base pairs (bp) of linker DNA \cite{Khorasan-2004}. The nucleosome consists of 147 bp DNA that is wrapped around the histone octamer in about 1.8 turns \cite{Richmond-2003}. This genomic organization and packaging plays a crucial role in regulating DNA accessibility and gene expression \cite{Cell}.  The arrangement of nucleosomes in chromatin fiber has been studied extensively in the past decade, via experimental approaches \cite{Schlick-2012,Kruithof-2009-Nat}. 

The chromatin structures beyond the nucleosomes have still come into question, although the structure of the 30 nm fiber {\it in vivo} has been debated in recent years \cite{Kruithof-2009-Nat,Tremethick-2007,Maeshima-2010}. The crystal structure of small array of four nucleosomes connected by 20 bp linker DNA reveals that next-neighbors of histone octamer are configured in a face-to-face manner and the linker DNA is straight \cite{Richmond-2005}. A similar structure has been seen using FRET (fluorescence resonance energy transfer) technique for the small array of three nucleosomes with 20 bp linker DNA \cite{Widom-2009}. Furthermore, recent measurements using cryogenic electron microscopy, studied the effect of protein H1 on the conformation of four nucleosomes connected by different length of linker DNA \cite{Song-2014}. In addition, in the different cell cycles other packaging conformation have be seen, for example in interphase and metaphase chromosomes \cite{Maeshima-2014}.

Computer simulations of the chromatin structure and dynamics have been developed over the past decade \cite{Boule-2015}, taking on different approaches that include coarse-grained \cite{Langowski-2011,Schlick-2014,Koslover-LocalGeometry} and all atom simulations \cite{Wong-2007,Ettig-2011}. Since an atomistic simulation for an array of nucleosomes with water molecules, salts and ions corresponding to the physiological conditions needs to consider a lot of particles, coarse-grained simulations and theoretical descriptions have become important. 
However, one may ask to what extent it is possible to understand the features of the problem using a simple elastic model. 

The experimental studies reveal the conformation of the DNA in the nucleosome core particle with very high precision \cite{Richmond-2003}. The conformational properties of the nucleosomal DNA have been studied theoretically and  the results are in a good agreement with the experimental data \cite{Farshid-PRL-2005,Maryam-NAR,Davood,Arman-Schiessel-EPJE}. In addition, the dynamics of the unwrapping and rewrapping of a nucleosome under force have been studied experimentally \cite{Mihardja-2006,Kruithof-2009,Tae-Hee-2012} and theoretically \cite{Kulic-PRL-2004,Sudhanshu-PNAS-2011,Laleh-BJ2012,Arya-BJ2012}. In these theories the DNA has been considered as an elastic rod wrapped around a cylinder, which is corresponding to a histone octamer. The success of the mentioned theories emphasizes the impact of the elastic description of the DNA in nucleosome.

According to the experimental observations, the length of the linker DNA varies in different situations \cite{Khorasan-2004} . These findings lead to an important question: What is the effect of the DNA elasticity in determination of the length of the linker DNA and the conformation of the arrays of the nucleosomes? Motivated by aforementioned problems, here we set out to consider a dinucleosome with a linker DNA that is flanked by a long DNA as shown in Fig. \ref{fig-schematic}. The aim of this paper is to show the effect of the linker DNA length and the DNA elasticity on the conformation of dinucleosomes. The rest of paper is organized as follows: In section II, we describe the model. After a general introduction of the model, we explain the way to find the energy of the dinucleosome structure considering proper boundary conditions. In section III, we present the results, and finally in section IV, we conclude the paper, while energy landscape for different forces and the DNA bending rigidities are explained in the Appendix.

\section*{II. Model}

\begin{figure}
\centering
\includegraphics[width=1\columnwidth]{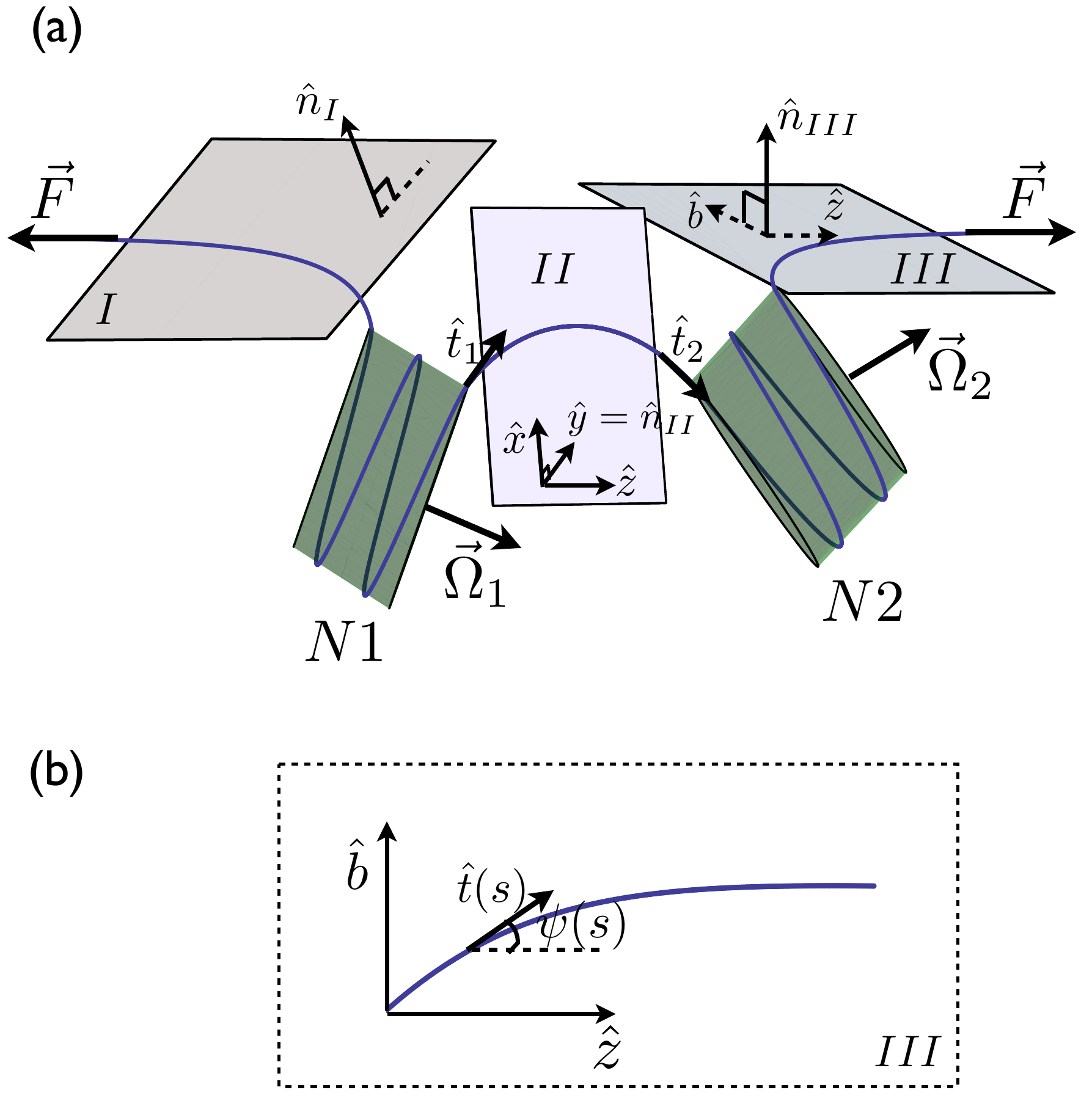}
\caption{ (Color online) (a) The schematic picture of a dinucleosome with two flanking DNA. The nucleosomes are denoted by $N1$ and $N2$. The flanking DNA are in the planes of I and III, whereas the linker DNA is in the plane of II. The orientation of the nucleosomes are shown by $\vec{\Omega}$'s. It is assumed that the force is applied in the $z$ axis and the Euler coordinate of $(x,y,z)$ is defined in the plane II, where the normal vector of that plane defines the $y$ axis. The normal vectors of the planes are shown by $\hat{n}$'s. (b) The schematic picture of the bent DNA in the plane III. The bending angle, $\psi(s)$, and the tangent unit vector, $\hat{t}(s)$, are shown in the figure. }
\label{fig-schematic}
\end{figure}

Here we consider a dinucleosome on a long DNA chain that is under tension by applying an extension force of $\vec{F} = F \hat{z}$ to its free ends. To study the conformation of the problem, we consider a simple model in which the DNA is represented as an elastic rod with the bending rigidity of $\kappa$. The molecule is parametrized by the arc length $s$, and $\hat{t}(s)$ denotes the unit tangent to the axis of the rod. As it is shown in Fig. \ref{fig-schematic}, there are five distinct regions of DNA: two nucleosomal DNA (shown by N1 and N2), two long flanking DNA (regions I and III), and one linker DNA (region II). Since the system can release the imposed twist, it is needed to consider only the bending energy of the deformed DNA. For an isotropic bent rod without any twist energy, the torque moment, $\vec{M}$, can be written as \cite{Landau}
\begin{eqnarray}
\vec{M} = \kappa \, \hat{t} \times \frac{d \hat{t}}{ds}. \label{eq:torque}
\end{eqnarray}
Using the force and torque balance equations one can find \cite{Landau}
\begin{eqnarray}
\kappa \, \hat{t} \times \frac{d^2 \hat{t}}{d s^2} = \vec{F} \times \hat{t}, \label{eq:equil}
\end{eqnarray}
We note that a bent isotropic rod without any imposed torsion, remains in a plane. For our case, each DNA regions are bent in a plane and hence $\hat{t}$, $\frac{d \hat{t}}{ds}$, $\frac{d^2 \hat{t}}{ds^2}$, and $\vec{F}$ are positioned in that plane, as shown in Fig. \ref{fig-schematic}. 

Therefore generally we will have three distinct plane corresponding to three bent DNA: two   flanking DNA and one linker DNA. We note that these three planes are not necessarily parallel, as can be seen in Fig. \ref{fig-schematic}(a).  

Since we have pure bending, the conformation of the isotropic rod can be determined by one angle. After defining $\psi(s)$ as the angle between the $\hat{t}$ and $\vec{F} = F \hat{z}$ (see Fig. \ref{fig-schematic}(b)), one can write the tangent vector as $\hat{t}(s) = \left( \sin \psi(s), \, 0 \, , \, \cos \psi(s) \right)$. Using the parametrization angle $\psi(s)$ and defining $\lambda \equiv \sqrt{\kappa/ (2 F)}$, the shape equation of the bent rod, Eq. (\ref{eq:equil}), can be written as
\begin{eqnarray}
2 \lambda^2 \, \ddot \psi = \sin \psi, \label{eq:dd-psi}
\end{eqnarray}
where we have used $\ddot \psi \equiv \frac{d^2 \psi}{d s^2}$. A first integration gives $\lambda^2 \dot{\psi}^2 = c - \cos \psi$, where $c$ is the integration constant and can be determined using boundary conditions. Now it is possible to read the coordinates of the rod at $s$ as
\begin{subequations}
\begin{eqnarray}
s(\psi) &=& \pm \lambda \int_{\psi_i}^\psi \frac{d \psi'}{\sqrt{c - \cos \psi'}}   \label{eq:s-psi}  \\ 
x (\psi ) &=&  \pm \lambda \int_{{\psi _i}}^{{\psi}} {\frac{{\sin \psi ' d\psi '}}{{\sqrt {c - \cos \psi '} }}}   \label{eq:x-psi}  \\
z (\psi ) &=&  \pm \lambda \int_{{\psi _i}}^{{\psi }} {\frac{{\cos \psi 'd\psi '}}{{\sqrt {c - \cos \psi '} }}},  \label{eq:z-psi} 
\end{eqnarray}
\end{subequations}
where $\psi_i$ denotes the angle of the rod with respect to the $z$ axis in the proper end of the rod. We can use the above equations for both flanking and linker DNA in our problem. We note that the constant $c$ can vary for the mentioned segments of the DNA. 

For the flanking DNA, one end of the molecule is absorbed to the histone octamer and the other end is under tension, see Fig. \ref{fig-schematic}. We assume that the absorbed end of the DNA can be considered as a clamped part. Now the problem is to determine the shape of a bent isotropic rod with one end clamped and the other end under a force $F$ parallel to the original direction of the rod. Since there is no imposed torque on the free end of the flanking DNA, the constant $c$ in the shape equation of the rod becomes 1 and we have $c_{flank} = 1$. In our problem, the free ends of the flanking DNAs are under tension and are along the $z$-axis, see Fig. \ref{fig-schematic}, and therefore one has
\begin{eqnarray}
\psi_i = 0. \quad {\rm flanking ~ DNA} \label{eq:psi_i_flanking}
\end{eqnarray}

For the linker DNA the situation is complex. First let us consider one of the nucleosomes. Since the DNA wraps around the histone octamer in 1.75 turns, the tangent vector of the linker DNA when it exits the nucleosome core is determined by the orientation of the nucleosome and the shape of the flanking DNA. Therefore, the orientations of the two nucleosomes, $\vec{\Omega_1}$ and $\vec{\Omega_2}$, and the conformation of the flanking DNAs, determine $\hat{t}_1$ and $\hat{t}_2$ uniquely, see Fig. \ref{fig-schematic}(a). But as we discussed above, $\hat{t}_1$ and $\hat{t}_2$ should be in the plane $II$. This criterion gives us the acceptable orientation of the nucleosomes $N1$ and $N2$. 
Furthermore, the torque and force balance conditions for each of the nucleosomes must be satisfied and this additional criterion gives us the acceptable conformation of the problem. 

\begin{figure}
\centering
\includegraphics[width=0.5\columnwidth]{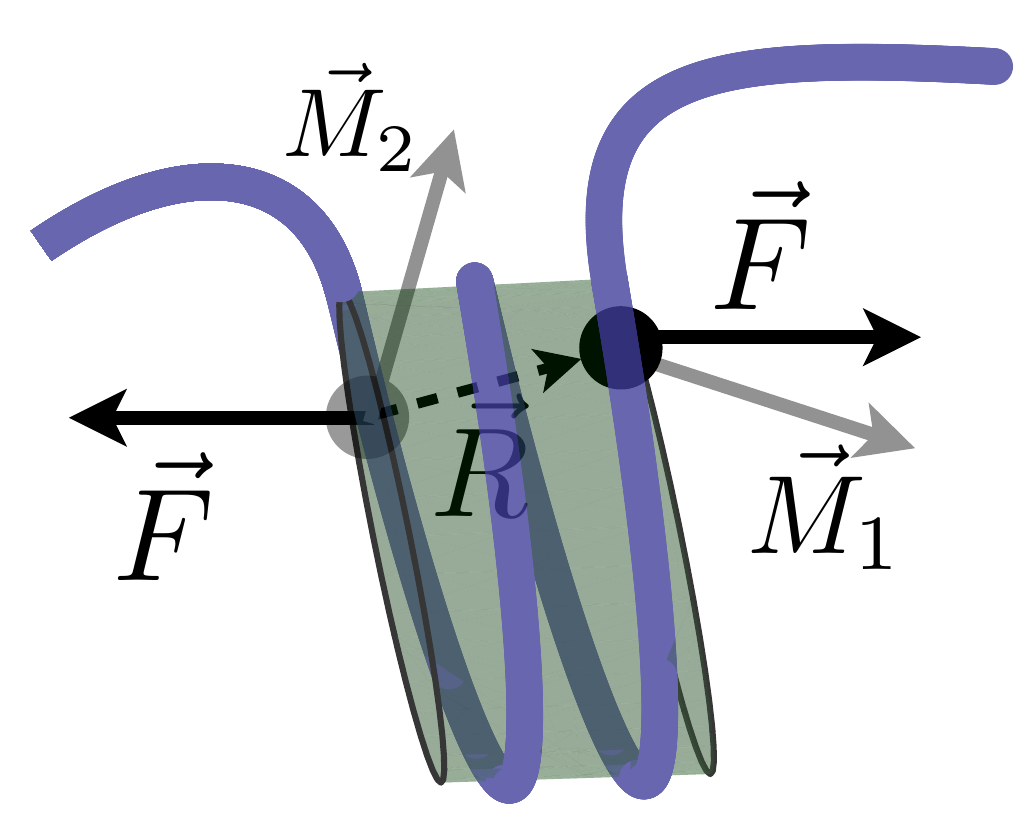}
\caption{ The forces and the torques acting on a nucleosome core. $\vec {F}$'s and $\vec{M}$'s denote the forces and torques acting on the nucleosome, respectively. These forces and torques act on the positions where the DNA leaves the nucleosome and shown by solid circles in the Fig., whereas $\vec{R}$ denotes the vector between the mentioned positions.}
\label{fig:nucleosome-M}
\end{figure}

Fig \ref{fig:nucleosome-M} shows the torques and forces that are applied on a nucleosome. According to torque balance equation one has
\begin{eqnarray}
\vec{M}_1+ \vec{M}_2 + \vec{R} \times \vec{F} = 0, \label{eq:torqe-balance}
\end{eqnarray}
where $\vec{R}$ is the vector between two points that the nucleosomal DNA leaves the histone octamer. It is worth mentioning that $\vec{M}_1$ and $\vec{M}_2$ can be found using Eq. (\ref{eq:torque}) and the orientation of the histone octamer determines $\vec{R}$. For the linker DNA, the constant $c$ should be determined in such a way that the torque balance condition for each nucleosome must be satisfied. Having the value of constant $c$, the boundary conditions and the constraints discussed above, one can find the shape of the rod using the Eqs. (\ref{eq:s-psi}) - (\ref{eq:z-psi}).

To derive the total energy of system, we need to account for the elastic energy of the deformed DNA,  the elastic energy of the nucleosomal DNA, the binding energy of the DNA at binding sites, and the effect of the external force on the free DNA portions. Generally, the total energy is a function of the orientation of the octamers, $\vec{\Omega}_i's$, the external force, $\vec{F}$, and the bending rigidity of the DNA, $\kappa$, and can be written as
\bea
E_{total}(\vec{\Omega}_1, \vec{\Omega}_2, \vec{F}, \kappa) = E_{nuc-DNA} + E_{free-DNA}, \label{eq:energy_total}
\eea
where $ E_{nuc-DNA}$ denotes the total energy of the nucleosomal DNA, and $E_{free-DNA}$ corresponds to the energy of the deformed free DNA portions.
First let us estimate the value of $ E_{nuc-DNA}$. This term has two contributions: (1) the binding energy of the DNA at binding sites, and (2) DNA deformation energy in the nucleosome structure. The two contributions  can be considered as an effective adsorption energy for the whole nucleosome. This energy is roughly $\sim - 40 k_BT$ \cite{Schiessel-2003,Laleh-BJ2012}, where the minus sign shows that the DNA prefers to wrap around the histone octamer in the physiological conditions. In this paper we deal with low force situations, i.e. $F < 2 pN$, where the DNA unwrapping from the octamer does not happen \cite{Mihardja-2006,Laleh-BJ2012} and the total wrapping energy may not change as a function of the external force and remains constant. The second term in the energy of Eq. (\ref{eq:energy_total}) has two contributions: (1) the bending energy of the DNA portions, and (2) the energy due to the presence of the external stretching force. This energy can be determined as 
\begin{eqnarray}
E = \frac{\kappa}{2} \int_{0}^L \dot{\psi}^2 ds - F \Delta z, \label{eq:energy}
\end{eqnarray}
where the first term corresponds to the bending energy, $E_{bend}$, and the $\Delta z \equiv z - z_0$ denotes the changes of the end-to-end distance of the DNA relative to the fixed $z_0$ at each $F$, whereas  is determined by $z = \int_0^L \cos \psi (s) ds $ \cite{Rob-Philips}. We set $z_0$ as the end-to-end distance of the DNA in the configuration of the problem corresponding to the lowest energy of the system. Therefore the energy of the problem corresponding to the global minimum energy can be considered as bending energy, $E_{bend}$ plus some constant value that corresponds to $E_{nuc-DNA}$. We note that by the way of choosing $z_0$, the term $F \Delta z$ is zero at the global minimum energy of the system. In the following section the results correspond to the lowest energy of the system, are shown.

\section*{III. Results}
\begin{figure}
\centering
\includegraphics[width=1\columnwidth]{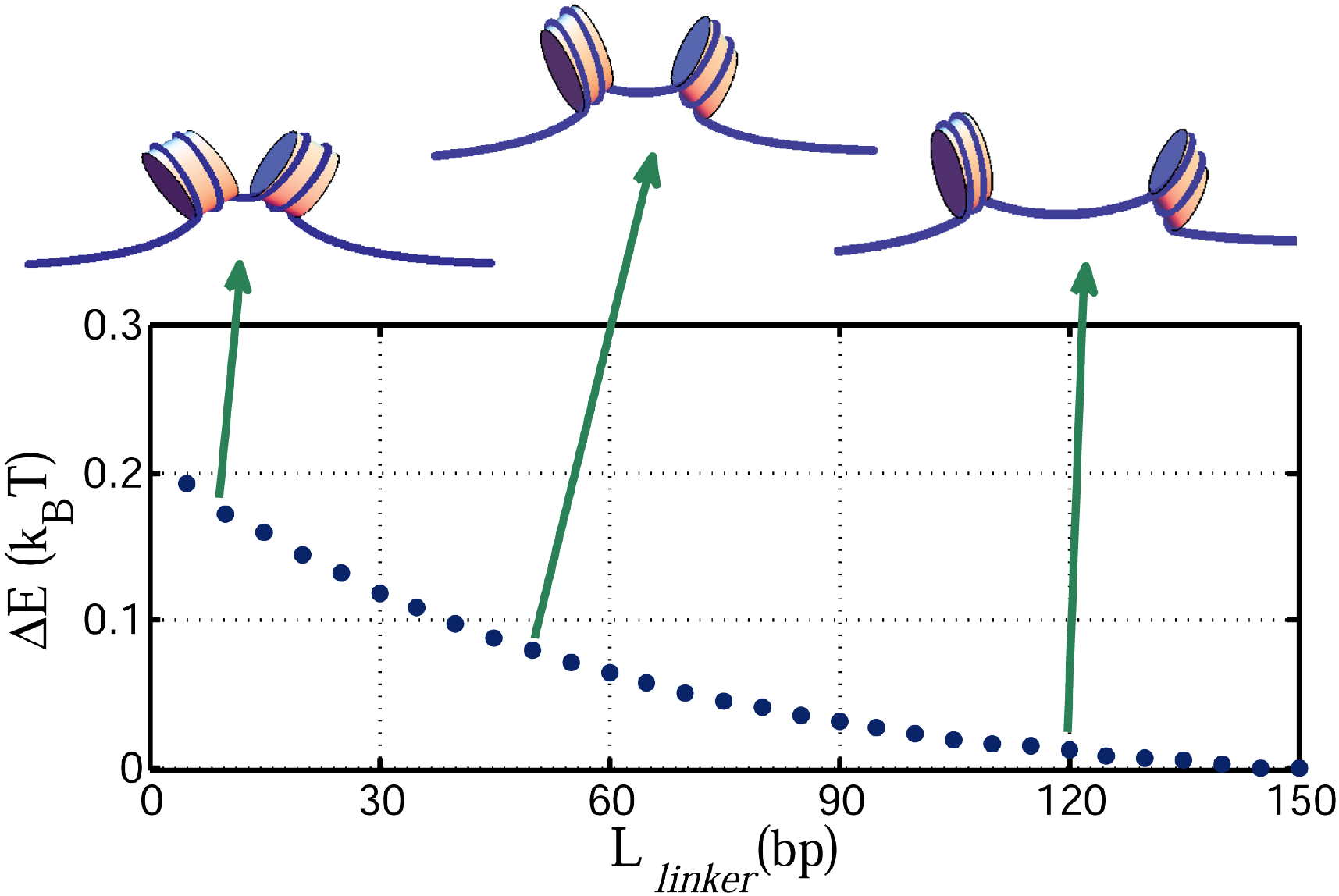}
\caption{(Color online) $\Delta E \equiv E_{min}(L_{linker}) - E_{min}(L_{linker} = 50 nm)$ versus the length of the linker DNA in the force of $F=0.5 $ pN. The conformations of the nucleosomes are shown for three different length of the linker DNA. }
\label{fig:E_L}
\end{figure}

In the nucleosome core particle, there are fourteen binding sites that the nucleosomal DNA is bound to the histone octamer via several hydrogen bonds between the histone proteins and the sugar-phosphate groups of the DNA backbone \cite{Davey-2002}. In these regions, the minor grooves of the DNA are positioned with the face to the nucleosome core proteins \cite{Luger-1997}. Since the intrinsic twist of B-DNA is $\sim 36^\circ$ per base pair, for the sake of simplicity, we only consider the situations where the linker DNA length is $5n $ bp with $n$ being an integer number. This guarantees that the excess twisting does not need to be considered in the system in order to keep the minor grooves face to the octamer.

As discussed in the previous section, each conformation of the dinucleosome may have a different energy and there is a conformation with the lowest energy, $E_{min}$ that can be calculated using Eq. (\ref{eq:energy_total}), which is corresponding to the optimized orientation of the nucleosomes. In Fig. \ref{fig:E_L}, the dependence of $\Delta E(L_{linker}) \equiv  E_{min}(L_{linker}) - E_{min} (L_{linkder} = 50 \, nm) $ on the length of the linker DNA in $F = 0.5$ pN is shown. 
Since the variation of the energy is smaller than $0.2 \, k_BT$, one can conclude that  the energy, more or less, does not depend on the length of the linker DNA. We can see the same behavior for other different forces, that are shown in the appendix. We note that in order to study the conformation of the dinucleosome structure, the external force should be sufficiently small to ensure that DNA is not unwrapped from the histone octamers.

\begin{figure}
\centering
\includegraphics[width=1\columnwidth]{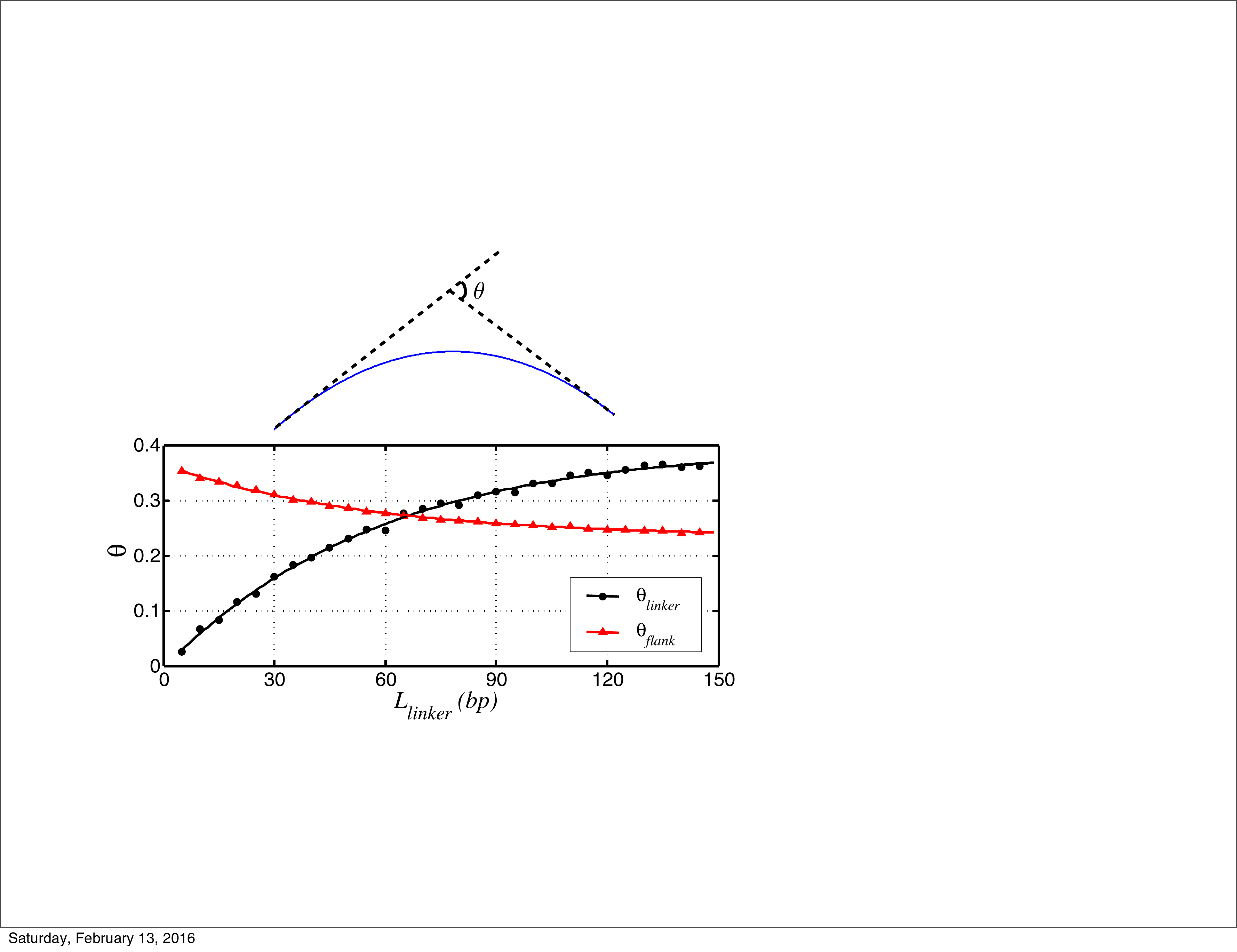}
\caption{(Color online) (top) The schematic picture of a bent rod for the overall bending angle of $\theta$.  (bottom)The overall bending angle of each segment of the DNA as a function of the length of the linker DNA, $L_{linker}$. The red triangles and the black circles are corresponding to the $\theta_{linker}$ and $\theta_{flank}$, respectively.   }
\label{fig:bending_L}
\end{figure}

In order to understand the uniform behavior in the energy landscape, we focus on the bending of the DNA segments. As seen in the dinucleosome problem, each segment of the DNA shown in Fig. \ref{fig-schematic} is bent. The bending of each segment is characterized by an overall bending angle, $\theta$. In Fig. \ref{fig:bending_L}, the overall bending angles of the flanking DNA, $\theta_{flank}$, and the linker DNA, $\theta_{linker}$, have been shown as a function of the linker DNA length. Since the bending of a longer rod is much easier than a shorter one, the longer the linker DNA, the larger the corresponding bending angle is, as can be seen in Fig. \ref{fig:bending_L}. Since the total length of the DNA has been considered constant in the problem, i.e. $L_{linker} + 2 L_{flank} = const.$, we can see that $\theta_{flank}$ becomes smaller for the longer linker DNA. These two behavior eventuates in an almost constant total energy as discussed in Fig. \ref{fig:E_L}. 

\begin{figure}
\centering
\includegraphics[width=1\columnwidth]{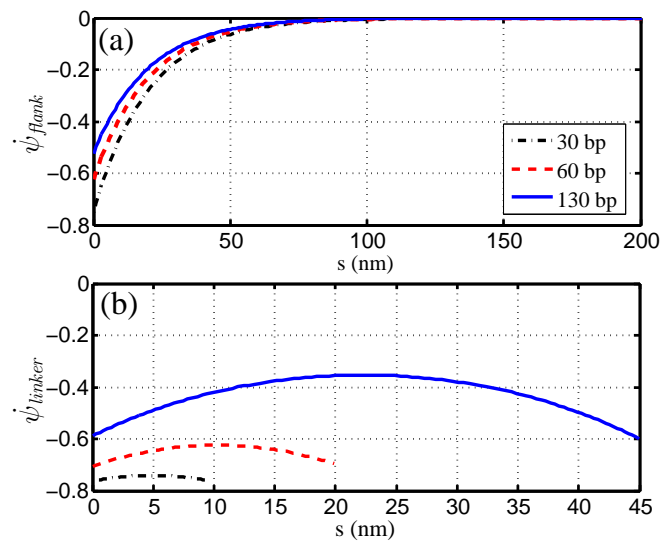}
\caption{(Color online) local $\dot{\psi}$ for $s$ for (a) the flanking DNA and (b) the linker DNA. The dash-dotted, dashed and solid lines are corresponding to the $L_{linker} = 30$ bp, $L_{linker} = 60$ bp, and $L_{linker} = 130$ bp, respectively. The external force has been considered as $F = 0.5$ pN.}
\label{fig:psidot-s}
\end{figure}

In figure \ref{fig:psidot-s}, the behavior of $\dot{\psi}$ is shown in terms of the contour length, $s$, for the flanking and the linker DNAs. As can be seen in the figure, $| \dot{\psi} | $ decreases with $s$ and becomes zero for large enough distance from the histone octamers. In the distance comparable to the persistence length of the rod, the variation on $\dot{\psi}$ is considerable. We note that for both two ends of the linker DNA, $\dot{\psi}$ should be the same, since the problem is symmetric about the middle of the DNA length. 
In addition the angle between the axis of cylinders (octamers) in the dinucleosome structure can be found as $\beta \equiv \cos^{-1}(\vec{\Omega}_1 \cdot \vec{\Omega}_2)$. This angle does not change with the linker DNA length and is approximately $\beta \simeq {70^\circ} \pm {6^\circ}$. It means the nucleosomes are almost perpendicular to each other and the relative spatial orientation of them is independent of the linker DNA length.

\section*{IV. Discussion}

In the above treatment, we have neglected details of the local DNA-histone interactions on the binding sites. Although there are some estimations on the DNA-histone interactions \cite{Schiessel-2003,Everaers-2015}, there is no reliable experimental measurements. Furthermore, the presence of ions in the solvent may have an affect on the local DNA-histone and histone-histone interactions \cite{Schiessel-2003}. 

These interactions can impose an effective moment in the problem, and according to the Eqs. (\ref{eq:torque}) and (\ref{eq:equil}), one can solve the problem using the same process as discussed in the text. Furthermore, in our model, the effect of twist is not considered. Since there is no external twisting torque on two ends of the flanking DNAs, any imposing twist can be washed out in the boundaries quite rapidly. Let us estimate the timescale for diffusing a twist kink through the DNA. We model the DNA as a cylinder of length $L$ and radius $r \simeq 1 $nm. The rotational drag coefficient of a cylinder around its axis is known to be $\mu = 4 \pi \eta r^2 L$, where $\eta$ is the solvent viscosity \cite{Tirado-1980}. Consequently, the ``rotational diffusion time'' of the cylinder is found as $t_{rot} \simeq \mu/k_BT = 4 \pi \eta r^2 L /k_BT$. For the flanking DNA we have $L \simeq 1 \mu$m and using a typical value for the viscosity, $\eta \simeq 10^{-3}$ Pa.s, we find the rotational time scale to be $t_{rot} \simeq 10^{-6}$s, which is comparable to the results of recent simulations \cite{Nam-Arya-2014}. This very small time scale indicates that any twisting in the problem can be washed out from the two free ends of the flanking DNA very rapidly. 

We note that in the biological conditions, histone tails can cause effective interactions between two neighboring nucleosomes. These interactions affect the relative orientation of nucleosomes and possibly the structure of linker DNAs \cite{Grigoryev-2009}. The mentioned interactions can be considered as an effective force and moment in the problem. Using the proposed model of this paper and Eqs. (\ref{eq:torque}) and (\ref{eq:equil}), one can incorporate these effective forces and moments and find out the structure of the linker DNA and the nucleosomes orientation.

We have neglected several other effects such as sequence inhomogeneity, rupture of DNA-histone bonds and nucleosome partial unwrapping and rewrapping due to thermal fluctuations. Different nucleotide sequences in the DNA structure may result in different values for $\kappa$. However one can define an effective bending rigidity for a given sequence of DNA and we expect that our model and its general results still hold. For this purpose, different values of  $\kappa$ has been considered in $F=0.5$ pN. As can be seen in the appendix, the energy landscape of dinucleosome does not change considerably. We note that according to Fig. \ref{fig:nucleosome-M} partial unwrapping of nucleosomes changes the acting points of the forces and torques and consequently the relative orientation of nucleosomes at the equilibrium state. It is worth mentioning that the force needed to open the first turn of the nucleosomal DNA is about 3 pN \cite{Mihardja-2006,Kruithof-2009,Laleh-BJ2012,Arya-BJ2012}, whereas the considered forces in this paper are smaller than 2 pN. 

In the places where the DNA leaves the nucleosomes shown in Fig. \ref{fig:nucleosome-M}, there are torques and forces that are applied on the nucleosomes and DNA. One can estimate the torque moment, $\vec{M}$, using the Eq. (\ref{eq:torque}). An estimate of the torque moment for the linker DNA length of 45 nm corresponding to Fig. \ref{fig:psidot-s} gives a value of $\sim 0.6 k_BT$, which is much smaller than the energy of the binding sites $\sim 5 k_BT$. Therefore the DNA-histone binding disturbance can be ignored in our model. In our model we do not consider fluctuations in the DNA flanking. This effect become important for forces smaller than the typical forces of $k_BT/\ell_p \simeq 0.08$ pN, where $\ell_p \simeq 50$ nm denotes the persistence length of the DNA. In our model we have studied the conformation and energy in the range of $0.5-2$ pN and therefore we study the problem in a pure energy landscape and neglect the thermal fluctuations.

In a recent measurement using Cryo-EM, Song {\it et al.} have determined the structure of 30 nm chromatin fiber with linker DNA lengths of 30 and 40 bp \cite{Song-2014}. They have observed that the overall structure of nucleosomes in the fiber with the linker DNAs have not been affected by increasing the length of the linker DNA, whereas the linker DNAs are almost straight. In another work, Schalch {\it et al.} have studied the tetra-nucleosomal conformation as a structural unit of 30 nm chromatin fiber with linker DNA length of 20 bp \cite{Richmond-2005}. Interestingly, they have found that there is a straight linker DNA in the tetra-nucleosome structure. Our model predicts that the dinucleosome structure does not depend on the length of the linker DNA, which may make us believe that the overall organization of the reconstituted chromatin fiber does not depend on the length of the linker DNA, which has been suggested in \cite{Richmond-2005,Song-2014}.

We finally propose the possible experimental setups for testing our findings. In order to see the dinucleosome structure, one can use a Cryo-electron microscopy similar to study of structure of 30 nm chromatin fiber \cite{Song-2014}. In addition, X-ray structure of the dinucleosome in a high precision measurement for different lengths of the DNA linker could reveal the orientation of the histone octamers with respect to each other as well as the conformation of the DNA linker.

In conclusion, we have also shown that by using a simple elastic model, the conformation of the dinucleosome system can be obtained. We have shown that in the force and moment balance conditions, the energy of the whole dinucleosome system does not depend strictly on the length of the linker DNA. Furthermore it has been shown that the orientation of the nucleosomes respect to each other does not vary in terms of the length of the linker DNA as well and they are almost perpendicular to each other, which is in good agreement with the observation \cite{Richmond-2005}. Our findings could shed some light on the structure of chromatin fiber and the dynamics of the nucleosome positioning in the physiological conditions.

\section*{Acknowledgments}
We are very grateful to Laleh Mollazadeh-Beidokhti for very fruitful discussions and comments. We thank Asal Atakhani and Maniya Maleki for very helpful comments on the manuscript. 

\section*{Appendix: Energy Landscape for Different Bending Rigidities and Forces}

In this appendix, we show the energy landscape of the dinucleosome structure in terms of the linker DNA length for different forces and bending rigidities. 

\begin{figure}[h!]
\centering
\includegraphics[width=1\columnwidth]{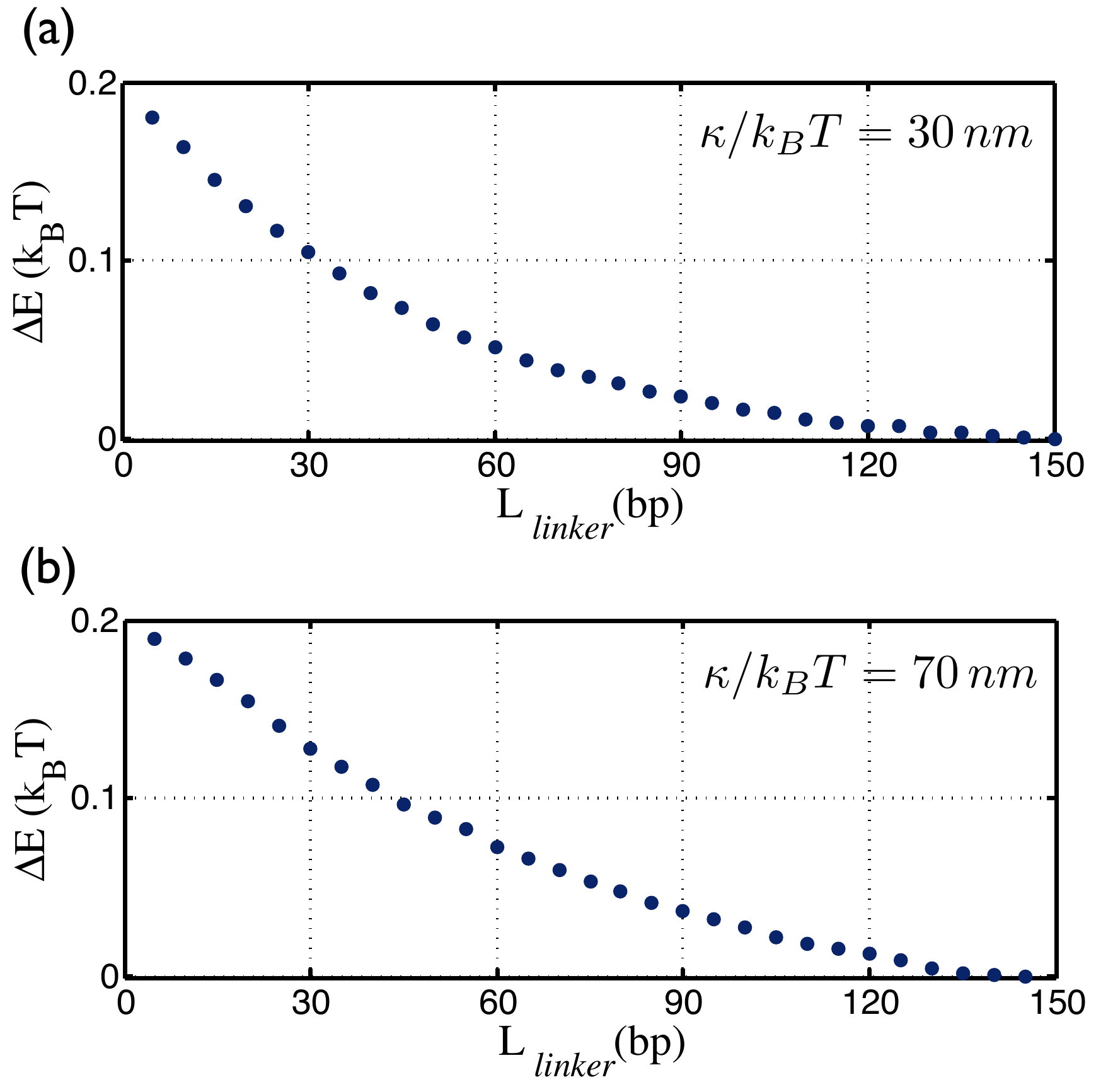}
\caption{$\Delta E \equiv E_{min}(L_{linker}) - E_{min}(L_{linker} = 50 nm)$ versus the length of the linker DNA for two different bending rigidity of the DNA, in the force of $F=0.5 $ pN. figures (a) and (b) are corresponding to $\kappa / k_BT = 30 nm$ and $\kappa / k_BT = 70 nm$, respectively. }
\label{fig:E_L_different_kappa}
\end{figure}

In Fig. \ref{fig:E_L_different_kappa}, $\Delta E$ in terms of the linker DNA length has been shown for two representative bending rigidities. The plot (a) corresponds to a ``soft'' DNA, whereas the plot (b) represents a ``hard'' DNA. As can be seen, the overall behavior that has been discussed in the main text still holds.

\begin{figure}[h!]
\centering
\includegraphics[width=1\columnwidth]{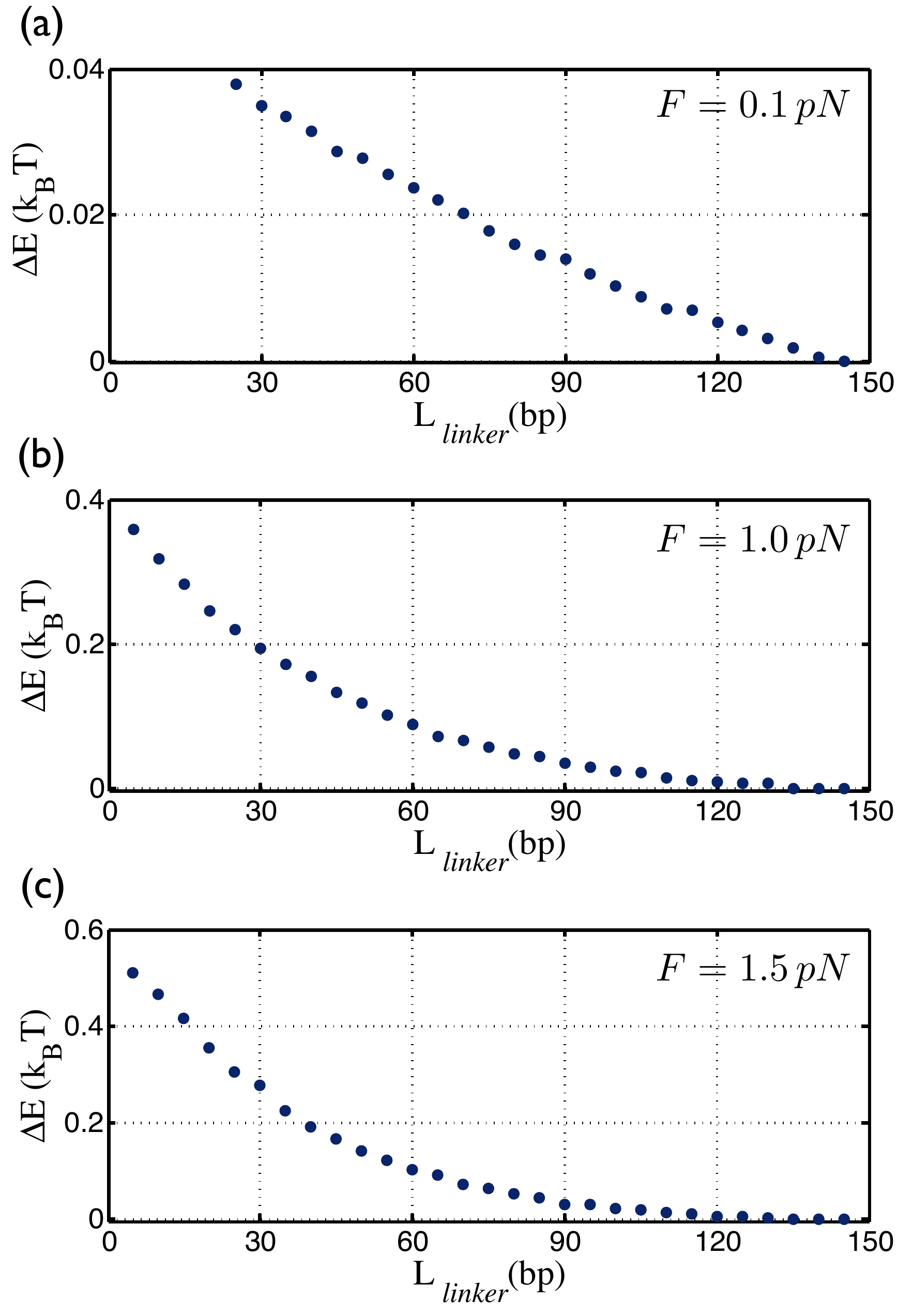}
\caption{$\Delta E$ in terms of the length of the linker DNA for different forces. (a) corresponds to $F = 0.1$ pN, (b) corresponds to $F = 1.0$ pN, and (c) corresponds to $F = 1.5 $ pN. The DNA bending rigidity is considered as $\kappa = 50 \, k_BT$.}
\label{fig:E_L_Force}
\end{figure}

In Fig. \ref{fig:E_L_Force}, $\Delta E$ in terms of the linker DNA length has been shown for three different forces. In plot (a), $F$ is considered as $0.1$ pN that corresponds to a low force limit, and (b) and (c) correspond to $F = 1.0$ pN and $F = 1.5$ pN, respectively.  As can be seen, increasing the length of the linker DNA does not affect the energy of the system.


\begin{thebibliography}{}

\bibitem{Khorasan-2004}
S. Khorasanizadeh, Cell {\bf 116}, 259 (2004).

\bibitem{Richmond-2003}
T. J. Richmond and C. A. Davey, Nature. {\bf 423}, 145 (2003).

\bibitem{Cell}
B. Alberts {\it et al.}, {\it Molecular Biology of the Cell} (Garland, New York, 2007), 5th ed.


\bibitem{Schlick-2012}
T. Schlick, J. Hayes, and S. Grigoryev, J. Biol. Chem. {\bf 17}, 5183 (2012).

\bibitem{Kruithof-2009-Nat}
M. Kruithof, F.T. Chien, A. Routh, C. Logie, D. Rhodes, and J. van Noort, Nat. Struct. Mol. Biol. {\bf 16}, 534 (2009).

\bibitem{Tremethick-2007}
D.J. Tremethick, Cell {\bf 128}, 651 (2007).

\bibitem{Maeshima-2010}
K. Maeshima, S. Hihara, and M. Eltsov, Curr. Opin. Cell Biol. {\bf 22}, 291 (2010).

\bibitem{Richmond-2005}
T. Schalch, S. Duda, D.F. Sargent, and T.J. Richmond, Nature {\bf 436}, 138 (2005).

\bibitem{Widom-2009}
M.G. Poirier, E. Oh, H.S. Tims, and J. Widom, Nature Struct. Mol. Biol. {\bf 16}, 938 (2009).

\bibitem{Song-2014}
F. Song {\it et al.}, Science {\bf 344}, 376 (2014). 

\bibitem{Maeshima-2014}
K. Maeshima, R. Imai, S. Tamura, and T. Nozaki, Chromosoma {\bf 123}, 225 (2014).

\bibitem{Boule-2015}
J. B. Boule, J. Mozziconacci, and C. Lavelle, J. Phys.: Cond. Matt., {\bf 27}, 033101 (2015).

\bibitem{Langowski-2011}
C. C. Fritsch, and J. Langowski, Chromosome Res. {\bf 19}, 63 (2011).

\bibitem{Schlick-2014}
R. Collepardo-Guevara, and T. Schlick, Proc. Natl. Acad. Sci. USA. {\bf 111}, 8061 (2014).

\bibitem{Koslover-LocalGeometry}
E. F. Koslover, C. J. Fuller, A. F. Straight, and A. J. Spakowitz, Biophys. J. {\bf 99}, 3941 (2010).

\bibitem{Wong-2007}
H. Wong, and J-M. Victor , and J. Mozziconacci, PLoS One {\bf 12}, e877 (2007).

\bibitem{Ettig-2011}
R. Ettig, N. Kepper, R. Stehr, G. Wedemann, and K. Rippe, Biophys. J. {\bf 101}, 1999 (2011).

\bibitem{Farshid-PRL-2005}
F. Mohammad-Rafiee, and R. Golestanian, Phys. Rev. Lett. {\bf 94}, 238102 (2005).

\bibitem{Maryam-NAR}
M. Ghorbani, and  F. Mohammad-Rafiee, Nucleic Acids Res. {\bf 39}, 1220 (2011).

\bibitem{Davood}
D. Norouzi, F. Mohammad-Rafiee, J. Biomol. Struct. Dyn {\bf 32}, 104 (2014).

\bibitem{Arman-Schiessel-EPJE}
A. Fathizadeh, A. B. Besya, M. R. Ejtehadi and H. Schiessel, Eur. Phys. J. E {\bf 36}, 21-1-10 (2013).

\bibitem{Mihardja-2006}
S. Mihardja, A. J. Spakowitz, Y. Zhang, and C. Bustamante, Proc. Natl. Acad. Sci. USA. {\bf 103}, 15871 (2006).

\bibitem{Kruithof-2009}
M. Kruithof, and J. van Noort, Biophys. J. {\bf 96}, 3708 (2009).

\bibitem{Tae-Hee-2012}
J. S. Choy, and T. H. Lee, Trends in biochemical sciences {\bf 37}, 425 (2012).

\bibitem{Kulic-PRL-2004}
I. M. Kuli$\acute{c}$, and H. Schiessel, Phys. Rev. Lett. {\bf 92}, 228101 (2004).

\bibitem{Sudhanshu-PNAS-2011}
B. Sudhanshu {\it et al.}, Proc. Natl. Acad. Sci. USA. {\bf 108}, 1885 (2011).

\bibitem{Laleh-BJ2012}
L. Mollazadeh-Beidokhti, F. Mohammad-Rafiee, and H. Schiessel, Biophys. J. {\bf 102}, 2235 (2012).

\bibitem{Arya-BJ2012}
I.V. Dobrovolskaia, and G. Arya, Biophys. J. {\bf 103}, 989 (2012).


\bibitem{Landau}
L.D. Landau, and E.M. Lifshitz. {\it Theory of Elasticity} (Pergamon Press Inc. New York, 1975), 3rd edition.

\bibitem{Rob-Philips}
R. Phillips, J. Kondev, J. Theriot. {\it Physical Biology of the Cell} (Garland Science Taylor \& Francis, New York, 2009), 1st edition. 

\bibitem{Davey-2002}
C.A. Davey, D.F. Sargent, K. Luger, A.W. Maeder, and T.J. Richmond, J. Mol. Biol., {\bf 319}, 1097 (2002).

\bibitem{Luger-1997}
K. Luger, A.W. Mader, R.K. Richmond, D.F. Sargent, and T.J. Richmond, Nature, {\bf 389}, 251 (1997).

\bibitem{Schiessel-2003}
H. Schiessel, J. Phys.: Condens. Matter {\bf 15}, R699 (2003).


\bibitem{Everaers-2015}
S. Meyer, and R. Everaers, J. Phys.: Condens. Matter {\bf 27}, 064101 (2015).


\bibitem{Tirado-1980}
M.M. Tirado, and J.G. de la Torre, J. Chem. Phys. {\bf 73}, 1986 (1980).
 
\bibitem{Nam-Arya-2014}
G.-M. Nam, and G. Arya, Nucleic Acids. Res. {\bf 42}, 9691 (2014).

 \bibitem{Grigoryev-2009}
S.A. Grigoryev {\it et al.}, Proc. Natl. Acad. Sci. USA. {\bf 106}, 13317 (2009).


\end{thebibliography}
\end{document}